\documentclass[journal,twoside,web]{ieeecolor} 
                                                    
\usepackage{generic}
\usepackage[utf8]{inputenc}
\usepackage{cite}
\usepackage[utf8]{inputenc}
\usepackage{cite}
\usepackage{qtree}
\usepackage{tikz}
\usepackage{amsmath,amssymb,amsfonts}
\usepackage{eufrak}
\usepackage{dsfont}
\usepackage{steinmetz}
\usepackage{algorithmicx}
\usepackage{algorithm}
\usepackage{algpseudocode}
\usepackage{graphicx}
\usepackage{amsmath,amssymb}
\usepackage{textcomp}
\usepackage{algpseudocode} 
\usepackage{mathtools}
\usepackage{nicefrac}
\usepackage{txfonts}
\usepackage{tikz}
\usepackage{dsfont}
\usepackage{graphicx}
\usepackage{amsmath,amssymb}
\usepackage{xcolor}
\usepackage{mathtools}
\usepackage{nicefrac}
\usepackage{txfonts}
\usepackage{todonotes}
\usepackage{cleveref}
\usepackage{subfig}
\usepackage[font=small]{caption}

\newcommand{\R}{\mathbb{R}}

\newcommand{\tr}{{\rm{tr}}}

\newcommand{\Q}{\mathcal{Q}}

\newcommand{\DQ}{\texttt{DQ}}
\newcommand{\cA}{c_A}
\newcommand{\cB}{c_B}

\renewcommand{\baselinestretch}{0.98}
\allowdisplaybreaks

\DeclareMathOperator{\argmin}{\arg\!\min}

\def\mb{\mathbb}
\def\mc{\mathcal}
\def\beq{\begin{equation*}}
\def\eeq{\end{equation*}}
\def\bql{\begin{equation}}
\def\eql{\end{equation}}
\def\bqn{\begin{eqnarray*}}
\def\eqn{\end{eqnarray*}}
\def\bnl{\begin{eqnarray}}
\def\enl{\end{eqnarray}}
\def\bna{\bql\begin{array}{rcl}}
\def\ena{\end{array}\eql}
\def\bnn{\beq\begin{array}{rcl}}
\def\enn{\end{array}\eeq}
\def\bma{\begin{bmatrix}}
\def\ema{\end{bmatrix}}
\def\bmx{\begin{matrix}}
\def\emx{\end{matrix}}
\def\ben{\begin{enumerate}}
\def\een{\end{enumerate}}
\def\bit{\begin{itemize}}
\def\eit{\end{itemize}}
\def\bei{\begin{itemize}}
\def\eei{\end{itemize}}
\def\bet{\begin{tabular}}
\def\eet{\end{tabular}}

\newtheorem{thm}{Theorem}

\newtheorem{df}{ Definition}

\newtheorem{rem}{ Remark}

\newtheorem{asm}{Assumption}

\title{\texttt{QSID-MPC}: Model Predictive Control with System Identification from Quantized Data}
\author{ Shahab Ataei, Dipankar Maity, and Debdipta Goswami
\thanks{S. Ataei is supported by OSU COE Strategic Research Initiative 2024 seed grant.}
\thanks{S. Ataei is with the Department of Electrical and Computer Engineering, The Ohio State University, Columbus,
OH, 43210, USA (e-mail:{ \tt ataei.3@osu.edu}).}
\thanks{D. Maity is with the Department of Electrical and Computer Engineering and an affiliated faculty of the North Carolina Battery Complexity, Autonomous Vehicle, and Electrification Research Center (BATT CAVE), University of North Carolina at Charlotte,  NC, 28223, USA (e-mail: {\tt {dmaity@charlotte.edu}}).
}
\thanks{D. Goswami is with the Department of Mechanical and Aerospace Engineering, The Ohio State University, Columbus,
OH, 43210, USA (e-mail:{ \tt goswami.78@osu.edu}).
}
}

\begin{document}
\maketitle
\thispagestyle{empty}

\begin{abstract}
Least-square system identification is widely used for data-driven model-predictive control (MPC) of unknown or partially known systems. This \textit{letter} investigates how the system identification and subsequent MPC is affected when the state and input data is quantized. Specifically, we examine the fundamental connection between model error and quantization resolution and how that affects the stability and boundedness of the MPC tracking error. Furthermore, we demonstrate that, with \textit{a sufficiently rich dataset}, the model error is bounded by a function of quantization resolution and the MPC tracking error is also \textit{ultimately bounded} similarly. The theory is validated through numerical experiments conducted on two different linear dynamical systems.

\end{abstract}


\section{Introduction} 
\IEEEPARstart{S}{ystem} identification is an essential component in the applications of control systems involving unknown or partially known dynamics.
%
For a linear time-invariant (LTI) system, the system matrices are commonly identified using a least-squares optimization approach, which relies on control and state data snapshots from the dynamical system.
It is well-understood that the quality of the system matrix estimates improves/degrades with an increase/decrease in the amount of data, as expected \cite{Arbabi2017}.
On the other hand, it is not clear how the quality of the data affects the estimation process, especially when the data undergo a quantization process. 
Furthermore, the effect of identification error due~to~data quality on the controller performance is also relatively~unexplored.

Existing research on data-driven system identification and control generally presumes that sufficient computational resources are available to implement identification and control algorithms capable of handling extensive datasets collected from dynamical system snapshots.
However, applying these data-intensive algorithms to resource-limited systems, such as low-powered, lightweight robotic applications \cite{Folkestad2022, Cleary2020}, may require quantization to meet hardware and other resource constraints. 
In fact, quantization naturally arises under communication and computation constraints, making it widely adopted in networked control systems, multi-agent systems, and cyber-physical systems.
Quantization can have severe consequences on control systems, to the extent that a stabilizable system becomes destabilized if the quantization word-length falls below a certain threshold \cite{nair2004stabilizability}.
%

Given that system identification is frequently the initial phase toward controlling systems with unknown dynamics, the impact of quantization on identification inherently propagates to controllers and state estimators, and ultimately influences the overall system performance.
The choice of the quantizer is also of particular significance, as it can affect the system's performance \cite{maity2021optimal, maity2023optimal}.

In this letter, we study the effects of \textit{dither quantization} \cite{gray1993dithered}---a widely adopted and highly effective quantization method in control, communication, and signal processing applications---on the data-driven MPC of unknown LTI systems, leading to a framework of Quantized System Identification for MPC, hereafter referred to as the \textbf{\texttt{QSID-MPC}} algorithm. Our prior works \cite{maity2024effect, maity2024EDMD, ataei2025koopman} delved into the effect of dither quantization on Koopman-based system identification method.
However, to the best of our knowledge, this is the first attempt to quantify the effect of quantized system identification on data-driven MPC performance.  

The main contributions of this \textit{letter} are as follows: (i) We establish explicit bounds on the identification error arising from quantized input and state data in the identification of LTI systems. In particular, we show that the model error depends on the quantization resolution $\epsilon$ as $O(\epsilon^2)$. (ii) We prove the \emph{uniform ultimate boundedness} of \textbf{\texttt{QSID-MPC}} tracking error. Furthermore, we show that this ultimate bound similarly depends on $\epsilon$ as $O(\epsilon^2)$. (iii) Our theory is validated through extensive experiments on two applications, testing various quantization resolutions.

The rest of the \textit{letter} is organized as follows: We define our problem statement in \Cref{sec:ProblemStatement} and analyze the effect~of dither quantization on system identification in~\Cref{sec:QuantizedDMD}, demonstra-ting the effect of quantization resolution on the~identification error. 
We provide the \textit{ultimate bound} on MPC~tracking error for the identified system in \Cref{sec:MPC}.
We discuss our observations from implementing \textbf{\texttt{QSID-MPC}} on two dynamical systems in \Cref{sec:Validation} and we provide some conclusions in \Cref{sec:conclusions}.

\textit{Notations:} Set of non-negative integers are denoted by $\mb{N}_0$. 
$(\cdot)^{\dagger}$ and $(\cdot)^\top$ denote the Moore--Penrose inverse and transpose of a matrix, respectively. 
$\|\cdot\|$ denotes a norm, where we use Euclidean norm for vectors and Frobenius norms for matrices. 
The Big-O notation is denoted by $O(\cdot)$.

\section{Problem Statement} \label{sec:ProblemStatement}
We consider a model predictive control problem for trajectory tracking of an \textit{unknown} linear system.
In particular, we consider a discrete-time linear time invariant system 
\begin{align} \label{eq:dynamics}
\begin{split}
    x_{t+1} = Ax_t + Bu_t,\quad & x_t\in\mc{M}\subset \mb{R}^n,\\ u_t\in\mc{U}\subset\mb{R}^m,\quad & t\in\mb{N}_0,
\end{split}
\end{align}
where both $A$ and $B$ matrices are unknown. 
We construct an MPC trajectory problem on an \textit{identified} system as follows:
\begin{align}\label{eq:LMPC}
        \begin{split}
        \underset{\{u_{k|t},~ x_{k+1|t}\}_{k=t}^{t+T_h-1}}{\text{  min   }} & \sum_{k=t}^{t+T_h-1} \left[(x_{k|t}-x^\text{ref}_{k|t
 })^{\top}Q(x_{k|t}-x^\text{ref}_{k|t
 }) + u_{k|t}^\top R u_{k|t}\right] \\ &+ (x_{t+T_h|t}-x^\text{ref}_{t+T_h|t
 })^{\top}Q_f(x_{t+T_h|t}-x^\text{ref}_{t+T_h|t
 })\\
        \text{subject to: } 
        & x_{k+1|t} = \hat A x_{k|t} + \hat B u_{k|t},\ k=t,\ldots,(t+T_h -1),\\
        & x_{k+1|t}\in \mc{M},~~ u_{k|t}\in\mc{U},~x_{t|t} = x_t,~x_{t+T_h|t}\in \mc{X}_f.
    \end{split}
\end{align}
where $\hat A$ and $\hat B$ are the identified system matrices, $T_h$ is the MPC horizon, and $\mc{X}_f$ is a terminal constraint required for the MPC stability. If no terminal constraint exists, then $\mc{X}_f = \mc{M}$.
We assume that $Q \succeq 0$ and $R \succ 0$ and known, as customary in standard linear quadratic (LQ) problems. The terminal cost matrix $Q_f$ is also added for proving the stability of the MPC as described in \Cref{sec:MPC}. The optimal value of the cost function depends on the current state $x_t$ and is denoted as $J^*(x_t)$.
The input computed from this MPC problem is applied to the original system \eqref{eq:dynamics}, and we expect the original system to track the reference trajectory $x^{\text{ref}}$. The resultant \textbf{\texttt{QSID-MPC}} algorithm is described in \Cref{Alg: MPC}.

In this problem, we investigate the scenario where the system identification of $({A}, {B})$ is performed with \textit{quantized data} and a least-square algorithm akin to the Dynamic Mode Decomposition method \cite{schmid2010, Williams2015}. 
While the least square problem---to be described soon---is able to \textit{accurately} identify the system under unquantized data, it fails to attain such perfect accuracy under quantized data. 
The objective of this \textit{letter} is to investigate the effect of the quantization on the system identification and, more generally, on the trajectory tracking MPC problem. 
Before proceeding further, let us assume the following on the original system \eqref{eq:dynamics}.

\begin{asm}\label{assm:controllability}
    $(A,B)$ is a controllable pair.
\end{asm}
This assumption is instrumental, as trajectory tracking may otherwise be unachievable, even with a perfectly identified system. 

\begin{rem}
    The effect of quantization on controller synthesis is a well studied problem \cite{fu2024tutorial}. However, quantization effects on system identification and its consequences on the following stages of controller synthesis have not yet been explored. 
    In this \textit{letter}, we will provide a fundamental relationship between the quantization resolution and the MPC performance and stability bounds by choosing $x^{\text{ref}}\equiv 0$.
\end{rem}

\begin{algorithm}[t]
\caption{\textbf{\texttt{QSID-MPC}}}\label{Alg: MPC}
\textbf{Input:} Reference trajectory $x^{\text{ref}}$, identified system matrices $[\hat A, \,\hat B]$, Cost weight matrices $Q$ and $R$, time horizon $T_h$\\
\textbf{Output:} Optimal control sequence $u^*_t,\,t\in \mb{N}_0$
\begin{algorithmic}[1]
\Procedure {Compute MPC}{$x^{\text{ref}}$, $\hat A$, $\hat B$, $Q$, $R$, $T_h$}
\State For each time $t\in \mb{N}_0$,
\State Set $x_{t|t} \gets x_t$;
\State Solve optimization \eqref{eq:LMPC};
\State $u^*_t \gets u^*_{t|t}$
\State Apply feedback control $u_t\gets u^*_t$ in \eqref{eq:dynamics} to get $x_{t+1}$;
\State Repeat for $t\gets t + 1$ until the end of the control task.
\EndProcedure
\end{algorithmic}
\end{algorithm}

\subsection{System Identification}

The system matrices $A$ and $B$ are identified from \textit{quantized} input and state data snapshots $\{\tilde{x}_t\}_{t=0}^T$ and $\{\tilde{u}\}_{t=0}^{T-1}$ as follows:
\begin{align}\label{Eq: optimization}
\begin{split}
    \hat G = [\hat A,\,\hat B]=& \underset{\mc{A} {\in \R^{n \times n}},\,\mc{B} \in {\R^{n \times m}}}{\argmin} \dfrac{1}{T}\|X^{+} - \mc{A} {X} - \mc{B} {U}\|^2 \\
    = & \underset{\mc{G} \in {\R^{n \times (n+m)}}}{\argmin} \dfrac{1}{T}\|X^{+} - \mc{G}\Psi\|^2,
    \end{split}
\end{align}
where 
\begin{align}
\label{eq:dataMatrix2}
\begin{split}
    {X} &= [\tilde x_0 ~ \hdots ~ \tilde x_{T-1}],\\
    X^{+} &= [\tilde x_1 ~ \hdots ~ \tilde x_{T}]\\
    {U} &= [\tilde u_0~\hdots~\tilde u_{T-1}], \text{ and} \\
    \Psi & = \begin{bmatrix}
        X^\top, 
        U^\top
    \end{bmatrix}^\top,
\end{split}
\end{align}
with $\tilde x_t$ and $\tilde u_t$ denoting the quantized versions of $x_t$ and $u_t$, respectively. 
Exact relationships between the unquantized and quantized variables will be provided soon after we discuss the quantization scheme.

\begin{rem} \label{rem:least-square}
    When the data is not quantized and enough amount of data is present, the least-square problem \eqref{Eq: optimization} has a unique solution $\mathcal{G}^* = [A, B]$. 
    For instance assume~$u_t = 0$ for all $t$, then the least square problem boils down~to $\min_{\mathcal{A}} \frac{1}{T}\|X^+ - \mathcal{A}X\|^2 = \min_{\mathcal{A}}  \frac{1}{T} \sum_{t=0}^{T-1}\|Ax_t - \mathcal{A}x_t\|^2 = \min_{\mathcal{A}}  \tr\left((A-\mathcal{A})\left(\frac{1}{T}\sum_{t=0}^{T-1}x_t x_t^\top\right) (A-\mathcal{A})^\top \right)$. 
    If enough data is accumulated to ensure $\frac{1}{T}\sum_{t=0}^{T-1}x_t x_t^\top \succ 0$, then $\mathcal{A}^* = A$ is the \textit{unique} optimal solution.
    Once matrix $A$ is identified, we may identify matrix $B$ in a similar fashion with enough data to ensure $\frac{1}{T}\sum_{t=0}^{T-1}u_t u_t^\top \succ 0$.
\end{rem}

While the least square algorithm \eqref{Eq: optimization} is quite effective in perfectly identifying $A$ and $B$, as shown in \Cref{rem:least-square},~quantization complicates things to a large extent. 
Our analysis first shows the effect of quantization on the identification errors $(A-\hat A)$ and $(B-\hat B)$, and then we provide a Lyapunov-based analysis to comment on the trajectory tracking and stability performances of the MPC controller developed using the identified system. 



\section{Effects of Dither Quantization on $(\hat{A}, \hat{B})$} \label{sec:QuantizedDMD}

In this work, we focus particularly on the \textit{dither quantization} (\DQ) scheme for two primary reasons: (1) Analytical tractability and (2) Superior performance of dither quantization compared to its non-dithered counterpart \cite{widrow1961statistical} under a carefully chosen dithering noise. 
In the \DQ~scheme, a noise is added to the signal prior to quantization, and then the same noise is subtracted from the decoded quantized signal. 
More precisely, let $\Q$ be a quantizer with range $[x_{\min}, x_{\max}] \subseteq \R$ and resolution $\epsilon$ such that for any $x \in \R$ 
\begin{align*}
    \Q(x) = \begin{cases} x_{\min} +\frac{\epsilon}{2} +
        \epsilon \left\lfloor\frac{x-x_{\min}}{\epsilon} \right\rfloor, \qquad & x \in [x_{\min}, x_{\max}], \\
        x_{\min} +\frac{\epsilon}{2}, & x< x_{\min}, \\
        x_{\min} +\frac{\epsilon}{2} +
        \epsilon \left\lfloor\frac{x_{\max}-x_{\min}}{\epsilon} \right\rfloor & x> x_{\max}.
    \end{cases}
\end{align*}
Quantizer $\Q$ requires $b = \lceil\log_2\frac{x_{\max}-x_{\min}}{\epsilon}\rceil$ bits to represent its quantized output.  
Under a $\DQ$ scheme, the quantized version of $x$ is given as 
\begin{align*}
    \tilde x = \Q(x+w) - w,
\end{align*}
where $w$ is the dithering noise. 
Consequently, the quantization error is defined as $e(x) = x-\tilde x = (x+w) - \Q(x+w)$. 

Notice that, both $\tilde x$ and $e$ are now random variables due to $w$, even though the original signal $x$ may not be. 
The primary characteristic of $\DQ$ facilitating both a tractable analysis in this letter \textit{and} its superior performance compared to alternative quantization methods \cite{widrow1961statistical} is the statistical independence of the quantization error $e(x)$ from the input signal $x$, achievable by selecting a suitable noise distribution.
While many such noise distributions exist (c.f., \cite{schuchman1964dither}), a uniform distribution $w \sim \mathcal{U}([-\frac{\Delta}{2}, \frac{\Delta}{2}])$ is one of the most popular one. 
We refer the readers to \cite{schuchman1964dither, gray1993dithered} for the historical context on $\DQ$ and \cite{ataei2025koopman} for a detailed discussion on $\DQ$ in system identification~context.

The main result of this section is summarized in the following theorem. 

\begin{thm}[large-data regime] \label{thm:large-data}
Suppose $T\to \infty$ and $ \lim_{T\to \infty} \frac{\Psi \Psi^\top}{T} $ is finite and positive definite, then 
    \begin{subequations} \label{eq:hat_delta}
    \begin{align}
        \hat{A} & = A - \epsilon^2 \Delta_A, \\
        \hat{B} & = B - \epsilon^2 \Delta_B,
    \end{align}    
    \end{subequations} 
    \begin{align*}
        \noindent \text{where }~[\Delta_A, \Delta_B] =  [A, B] \left( 12 \lim_{T\to \infty}\frac{\Psi_{\rm uqz} (\Psi_{\rm uqz})^\top}{T} + \epsilon^2 I \right)^{-1},
    \end{align*}
    with $\Psi_{\rm uqz}$ being the data matrix  $\Psi$ constructed from unquantized data.
\end{thm} 

\begin{proof}
    The proof follows directly from \cite[Theorem~3]{ataei2025koopman}, which proves that $[\hat{A}, \hat{B}]$, as defined in \eqref{Eq: optimization}, can be written as 
    \begin{align*}
        [\hat A, \hat B] = [A_{\rm uqz}, B_{\rm uqz}] - \frac{\epsilon^2}{12} [A_{\rm uqz}, B_{\rm uqz}]\! \left(  \lim_{T\to \infty}\!\!\frac{\Psi_{\rm uqz} \Psi_{\rm uqz}^\top}{T} + \frac{\epsilon^2}{12} I \right)^{-1}\!\!\!\!,
    \end{align*}
    where $[A_{\rm uqz}, B_{\rm uqz}]$ is the identified system (i.e., solution to \eqref{Eq: optimization}) with \textit{unquantized} data.
    In \Cref{rem:least-square}, we show one can accurately identify $[A,B]$ with unquatized data, leading to the relationship
    \begin{align*}
        [\hat A, \hat B] = [A, B] - \epsilon^2 [A, B] \left( 12 \lim_{T\to \infty}\frac{\Psi_{\rm uqz} \Psi_{\rm uqz}^\top}{T} + \epsilon^2 I \right)^{-1}\!\!,
    \end{align*}
    which proves this theorem. 
\end{proof}
Let $\cA = \|\Delta_A\|$ and $\cB = \|\Delta_B\|$ for the subsequent~analysis.

\Cref{thm:large-data} shows how the quantization resolution affects the identification process, providing a fundamental connection between data quality and identification accuracy. 
Another interesting observation is that $[\hat A, \hat B]$, as computed in \eqref{Eq: optimization}, would be a random matrix due to the dithering noise introduced via quantization (the data matrix $\Psi$ is dithering noise dependent). 
However, as shown in \eqref{eq:hat_delta}, it appears that $[\hat A, \hat B]$ is a deterministic quantity since neither $[A,B]$ nor $[\Delta_A, \Delta_B]$ depends on the dithering noise. 
This is because the noises `average out' in the large data regime: a phenomenon that is fundamental to system identification under dither quantization in a large data regime, as originally shown in \cite{maity2024effect} and then later in \cite{maity2024EDMD, ataei2025koopman}.




Although \textit{large-data regime} provides a clear relationship between the model mismatch and the quantization error, we next discuss some \textit{fine-data regime} results, as numerical algorithms will have finite amount of data.


\begin{thm}[finite-data regime] \label{thm:K_epsilon}
    Let $\Psi$ and $\Psi_{\rm uqz}$ be of full row rank.
    Then, $\exists \ \mathcal{G}_\epsilon$ such that $\|\mathcal G_\epsilon\| = O(\epsilon)$ and 
    \begin{align}
    [\hat A, \hat B ] = [A, B] + \mathcal G_\epsilon.
    \end{align}
\end{thm}

\begin{proof}
    The closed form solution to the least-square problem in \eqref{Eq: optimization} with quantized data  is 
\begin{align} \label{eq:KDt_solution}
    [\hat A, \hat B] = X^{+}  \Psi^\top \big(  \Psi  \Psi^\top\big)^{-1}, 
\end{align}
whereas that for the unquantized case is $G_{\rm uqz} =  X_{\rm uqz}^{+}  \Psi_{\rm uqz}^\top \big(  \Psi_{\rm uqz}  \Psi_{\rm uqz}^\top\big)^{-1}$.

Let $e^x_t = x_t - \tilde{x}_t$ and $e^u_t = u_t - \tilde{u}_t$ denote the quantization error of the state and control, respectively, at time $t$. 
Let us further define the error matrices $E^x = [e_0^x,e_1^x,\ldots,e_{T-1}^x]$, ${E^x}^+ = [e_1^x,e_2^x,\ldots,e_{T}^x]$, $E^u = [e_0^u,e_1^u,\ldots,e_{T-1}^u]$, and $E^\Psi = [{E^x}^\top, {E^u}^\top]^\top$. 
Consequently, $X = X_{uqz}+E^x$, $X^+ = X^+_{uqz} + {E^x}^+$, $U= U_{uqz} + E^u$ and $\Psi = \Psi_{uqz} + E^\Psi$, and finally, 
\begin{align*}
    [\hat A, \hat B] =& (X^+_{uqz} + {E^x}^+)  (\Psi_{uqz} + E^\Psi)^\top \big(  (\Psi_{uqz} + E^\Psi)  (\Psi_{uqz} + E^\Psi)^\top\big)^{-1} \\
    = & G_{\rm uqz} - G_{\rm uqz}K + L
\end{align*}
where $K = \left(\Psi_{\rm uqz}  \Psi_{\rm uqz}^\top M_\epsilon^{-1} + I \right)^{-1}$ and $L = N_\epsilon(\Psi \Psi^\top)^{-1}$ and 
\begin{align*}
    M_\epsilon &= E^\Psi \Psi^\top + \Psi {E^\Psi}^\top + E^\psi{E^\psi}^\top, \\
    N_\epsilon & = {E^x}^+ \Psi_{\rm uqz}^\top + X^+ {E^\Psi}^\top + {E^x}^+ {E^\Psi}^\top.
\end{align*}
This completes the proof with $\mathcal{G}_\epsilon = - G_{\rm uqz}K + L$. 
Since the quantization error is in the range $[-\frac{\epsilon}{2}, \frac{\epsilon}{2}]$, we conclude that $\mathcal{G}_\epsilon = O(\epsilon)$. 
\end{proof}

\begin{rem}
    In both large- and finite-data regimes, the identification error diminishes as $\epsilon \to 0$. 
    Furthermore, since the quantization resolution $\epsilon$ is proportional to $2^{-b}$, where $b$ is the word-length of the quantizer, the system identification errors decays exponentially with the word-length---a phenomenon hat will be evident in all simulation results. 
\end{rem}


\section{Stability of Model Predictive Control (MPC) under Identification Error}\label{sec:MPC}
In this section, we seek to prove \emph{uniform ultimate boundedness} or \emph{practical asymptotic stability} of the LTI system \eqref{eq:dynamics} under the \textbf{\texttt{QSID-MPC}} framework of \Cref{Alg: MPC}. To this end, we first state an assumption on the identified system $[\hat A,\,\hat B]$.
\begin{asm} \label{assm:controllability_identified}
 $[\hat A,\,\hat B]$ form a controllable pair.  
\end{asm}
Our goal is to show that, under Algorithm \ref{Alg: MPC}, $x_t$ tracks the reference $x^{\text{ref}}$ with a \emph{uniformly ultimately bounded} error even in the presence of model error $[\epsilon^2\Delta_A,\,\epsilon^2\Delta_B]$.
The MPC problem \eqref{eq:LMPC} focuses on tracking an arbitrary time-varying trajectory $x^{\text{ref}}$. However, the \emph{ultimate boundedness} of the tracking error can be established only if $x^{\text{ref}}$ is constant or asymptotically converging to a constant. Without loss of generality, we assume $x^{\text{ref}}\equiv 0$ in this section. With the origin as a reference, we now define the \emph{uniform ultimate boundedness} of the MPC \Cref{Alg: MPC}.
\begin{df}
Let the MPC optimization problem \eqref{eq:LMPC} be feasible for all $x_t\in \mc{M}$. Then the solution of the system \eqref{eq:dynamics} under MPC policy with $x^{\text{ref}}\equiv 0$ from \cref{Alg: MPC} is called \emph{uniformly ultimately bounded with an ultimate bound} $\delta$ if there exists a class $\mc{KL}$ function  $\beta$, and for every initial state $x_0$ satisfying $\|x_0\| \leq r$ with $r >\delta $, there exists $T\geq 0$ such that
\[\|x_t\| \leq \beta (\|x_0\|, t),~\forall~t\in\{0,\ldots, T\}~\text{and}~
\|x_t\| \leq \delta,~\forall t> T.\]
\end{df}
This is also called \emph{practical asymptotic stability} \cite{Grune2017}.
\begin{asm}\label{assm:cost}
    The cost weight matrices are positive \emph{definite}, i.e, $Q, R \succ 0$. More specifically $x^\top Q x + u^\top R u \geq \alpha_1 \|x\|^2 + \alpha_2 \|u\|^2$.
\end{asm}
Note that, this is stronger than $Q\succeq 0$ prevalent in standard LQ problems and required for \emph{ultimate boundedness} of the error.
\begin{asm}\label{assm:feas}
    Feasibility--There exists a feasible control policy $\mu(\cdot)$ for which $\mc{X}_f$ is invariant, i.e., $\mu(\hat x) \in \mc{U}~\forall~\hat x\in\mc{X}_f$ and $\hat A \hat x + \hat B \mu(\hat x) \in \mc{X}_f$.
\end{asm}
\begin{rem}
    Assumption \ref{assm:feas} is trivially satisfied if $[\hat A, \hat B]$ is a controllable pair and $\mc{X}_f$ can be an arbitrary level set of any valid Lyapunov function under stabilizing control $\mu(x) = -Kx$ such that $\hat A- \hat BK$ has all eigenvalues within the unit circle.
\end{rem}

\begin{asm}\label{assm:Lyap}
    Let $V_f(x) \triangleq x^\top Q_f x$ and $l(x,u) \triangleq x^\top Q x + u^\top R u$, where $Q,~R,$ and $Q_f$ are as defined in \eqref{eq:LMPC}. $V_f(x)$ is a discrete-time Lyapunov function for $\hat x_{t+1} = \hat A \hat x_t + \hat B u_t$ with $u_t=\mu(\hat x_t)$ as defined in \Cref{assm:feas} within $\mc{X}_f$. Furthermore,
    $ V_f(\hat x_{t+1}) - V_f(\hat x_t) \leq -l(\hat x_t, \mu(\hat x_t))$ for all $x_t \in \mc{X}_f$.
\end{asm}

Now we are ready to state the central result concerning the uniform ultimate boundedness of the system dynamics \eqref{eq:dynamics} under MPC \cref{Alg: MPC}.

\begin{thm}
    The system \eqref{eq:dynamics} under a control policy computed via \Cref{Alg: MPC} is uniformly ultimately bounded with an ultimate bound $\delta(\epsilon)$ as $t\rightarrow \infty$.
\end{thm}

\begin{proof}
    Let us choose a Lyapunov function $V(x_t) \triangleq J^*(x_t)$, i.e., the optimal cost of \eqref{eq:LMPC} for computed control sequence $\{u^*_{t|t},\ldots,u^*_{t+T_h-1|t}\}$. Now, shifting the control sequence by one step and using the feasible control from \Cref{assm:feas}, we get a new sequence $\{u^*_{t+1|t},\ldots,u^*_{t+T_h-1|t}, \mu(x^*_{t+T_h|t})\}$. Since this control sequence is not optimal for \eqref{eq:LMPC} starting with $\hat{x}_{t+1}\triangleq x_{t+1|t}$, we can create a suboptimal cost function 
    \[J'(\hat{x}_{t+1}) = \sum\limits_{k=t+1}^{t+T_h} l(x^*_{k|t},u^*_{k|t}) + V_f( x_{t+T_h+1|t}),\] where $u^*_{t+T_h|t} \triangleq \mu(x^*_{t+T_h|t})$ and ${x_{t+T_h+1|t}} = \hat Ax^*_{t+T_h|t} + \hat Bu^*_{t+T_h|t}$. However, it is evident that the optimal cost of \eqref{eq:LMPC} $J^*(\hat{x}_{t+1})$ starting from $\hat{x}_{t+1} = x_{t+1|t}$ should satisfy
    \begin{align*}
    \begin{split}
        &J^*(\hat x_{t+1}) \leq  J'(\hat x_{t+1})\\
        &= J^*(x_t) - l(x^*_t, u^*_t) - V_f(x^*_{t+T_h|t}) \\
        &+V_f(x_{t+T_h+1|t}) + l(x^*_{t+T_h},\mu(x^*_{t+T_h|t}))
    \end{split}
    \end{align*}
    From \Cref{assm:Lyap}, we know $-V_f(x^*_{t+T_h|t}) 
        +V_f(x_{t+T_h+1|t}) + l(x^*_{t+T_h},\mu(x^*_{t+T_h|t}))\leq 0$, yielding 
    \begin{align}\label{eq:lyap_descent}
        V(\hat x_{t+1}) - V(x_t) \leq\! - l(x^*_t, u^*_t) = - l(x_t, u_t),\forall~x_t \neq 0,\forall~t\geq 0.
    \end{align}
    Now, the true next state is $x_{t+1} = \hat x_{t+1} + d(x_t,u_t)$, where $d(x_t,u_t)$ $\triangleq \epsilon^2(\Delta_A x_t + \Delta_B u_t)$ and satisfies $\|d(x_t,u_t)\|\leq \epsilon^2(c_A \|x_t\| + c_B\|u_t\|)$. Since $V(x_t)$ is the finite-horizon linear-quadratic optimal cost as evident from \eqref{eq:LMPC}, $V(x_t) = x_t^\top P x_t$ where $P=P_0$ is the solution of the Riccati equation 
    \[P_{k-1} = \hat A^\top P_k \hat A - \hat A^\top P_k \hat B (R+\hat B^\top P_k \hat B)^{-1}\hat B^\top P_k \hat A + Q,~P_{T_h} = Q_f.\]
    Hence, $\lambda_{\text{min}}(P)\|x_t\|^2 \leq V(x_t) \leq \lambda_{\text{max}}(P)\|x_t\|^2$. Also, choosing a sufficiently large $\rho>0$ will ensure $x_t\in B_{\rho}(0)$ for all $t>0$ and $V$ is \emph{uniformly continuous} on $B_{\rho}(0)$ by construction. Hence, there exists $\omega_v\in\mb{R}_+$ such that
    \begin{align}
        \begin{split}
            &V(x_{t+1}) = V(\hat x_{t+1} + d(x_t,u_t)) \leq V(\hat x_{t+1}) + \omega_v\|d(x_t,u_t)\| \\ 
            &\leq V(\hat x_{t+1}) + \omega_v\epsilon^2(c_A \|x_t\| + c_B\|u_t\|)\\
            &\leq V(\hat x_{t+1}) + \dfrac{\alpha_1}{2}\|x_t\|^2 + \dfrac{\omega_v^2\epsilon^4c_A^2}{2\alpha_1} + \dfrac{\alpha_2}{2}\|u_t\|^2 + \dfrac{\omega_v^2\epsilon^4c_B^2}{2\alpha_2}\\
            &=V(\hat x_{t+1})+\left( \dfrac{\alpha_1}{2}\|x_t\|^2 + \dfrac{\alpha_2}{2}\|u_t\|^2 \right) + \left( \dfrac{\omega_v^2\epsilon^4c_A^2}{2\alpha_1} + \dfrac{\omega_v^2\epsilon^4c_B^2}{2\alpha_2} \right),
        \end{split} \label{eq:new_inequality}
    \end{align}
    where the third inequality follows from Young's inequality.
    Substituting \eqref{eq:lyap_descent} in \eqref{eq:new_inequality}  yields,
    \begin{align}\label{eq:lyap_true}
            V(x_{t+1}) - V(x_t) &\leq - \left( \dfrac{\alpha_1}{2}\|x_t^2\| + \dfrac{\alpha_2}{2}\|u_t^2\| \right) +  \left( \dfrac{\omega_v^2\epsilon^4c_A^2}{2\alpha_1} + \dfrac{\omega_v^2\epsilon^4c_B^2}{2\alpha_2} \right)\nonumber\\
            &\leq - \dfrac{\alpha_1}{2}\|x_t^2\| +  C(\epsilon),
    \end{align}
    where we have used $\hat x_t = x_t$ and  $l(x,u)\!=\!x^\top\! Q x + u^\top\! R u \ge $ $ \alpha_1 \|x\|^2 + \alpha_2 \|u\|^2$ from \Cref{assm:cost} and defined the $O(\epsilon^4)$ function $C(\epsilon) \triangleq \omega_v^2\epsilon^4 \left( \dfrac{c_A^2}{2\alpha_1} + \dfrac{c_B^2}{2\alpha_2} \right)\geq 0$. Finally, \eqref{eq:lyap_true} can be rewritten as 
    \begin{align}
    \begin{split}
        V(x_{t+1}) & - V(x_t) \leq -(1-\theta)\dfrac{\alpha_1}{2}\|x_t\|^2 - \theta \dfrac{\alpha_1}{2}\|x_t\|^2 + C(\epsilon)\\
        &\leq -(1-\theta)\dfrac{\alpha_1}{2}\|x_t\|^2~\text{when}~\|x_t\|\geq \sqrt{\dfrac{2C(\epsilon)}{\theta\alpha_1}},
    \end{split}
    \end{align}
    where $\theta\in(0,1)$. From Theorem 2.6 in \cite{Cruz-Hernandez1999} or Theorem 4.18 in \cite{Khalil2015}, we conclude that the system \eqref{eq:dynamics} is \emph{uniformly ultimately bounded} by $\delta(\epsilon)\triangleq \sqrt{\dfrac{\lambda_{\text{max}}(P)}{\lambda_{\text{min}}(P)}\left(\dfrac{2C(\epsilon)}{\theta\alpha_1}\right)}$ which is an $O(\epsilon^2)$ function. This completes the proof.
\end{proof}

\section{Numerical Examples} \label{sec:Validation}
In this section, we present numerical examples to illustrate the theoretical results on the effect of quantization-induced identification error on MPC.
We demonstrate how varying the quantization resolution influences both system identification error and the resulting MPC regulation behavior for two systems:
a DC Motor with Load and control of Boeing 747 longitudinal flight modes.

\begin{figure*}
\centering 
\subfloat[]{\includegraphics[trim=1.9cm 7cm 2cm 7cm, clip=true, width=0.33\textwidth]{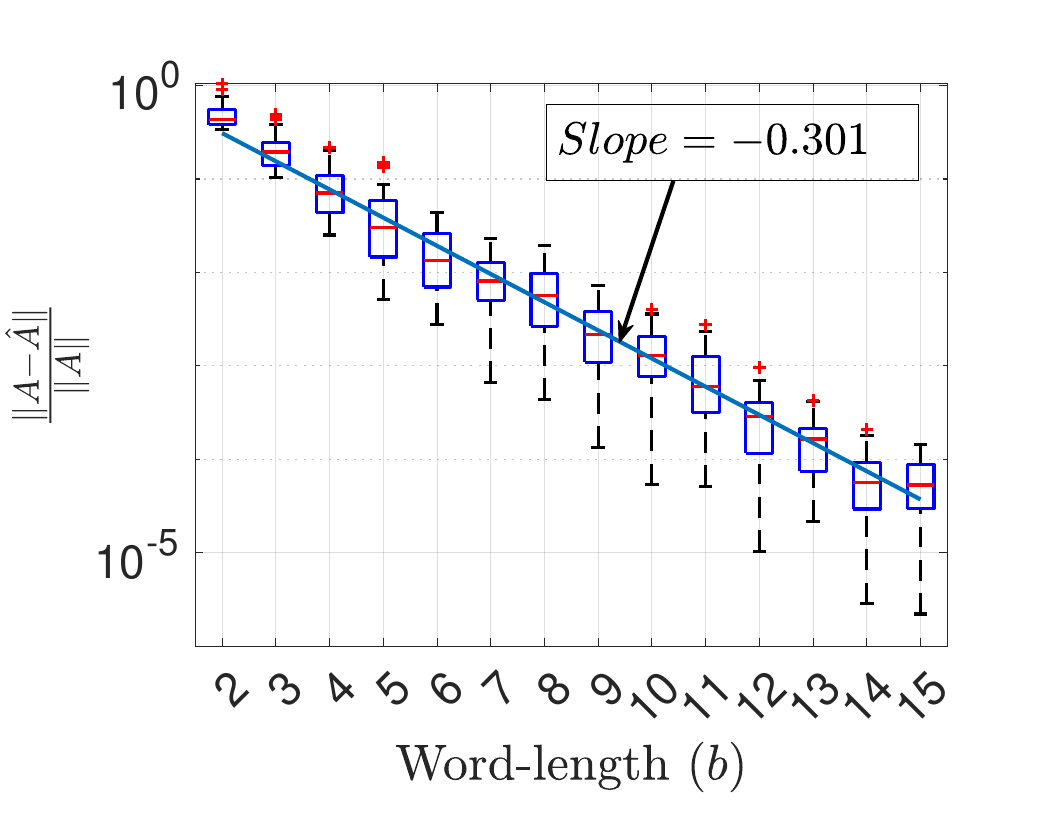}}
\subfloat[]{\includegraphics[trim=1.9cm 7cm 2cm 7cm, clip=true, width=0.33\textwidth]{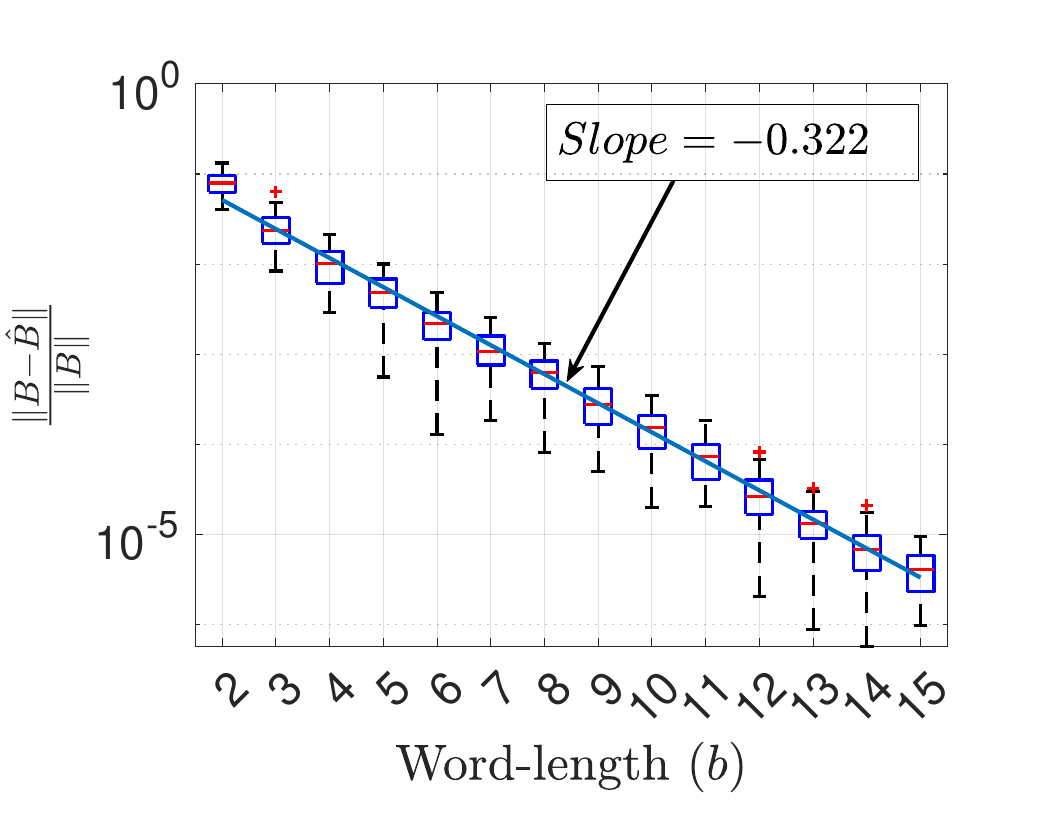}}
\subfloat[]{\includegraphics[trim=2cm 7cm 2cm 7cm, clip=true, width=0.33\textwidth]{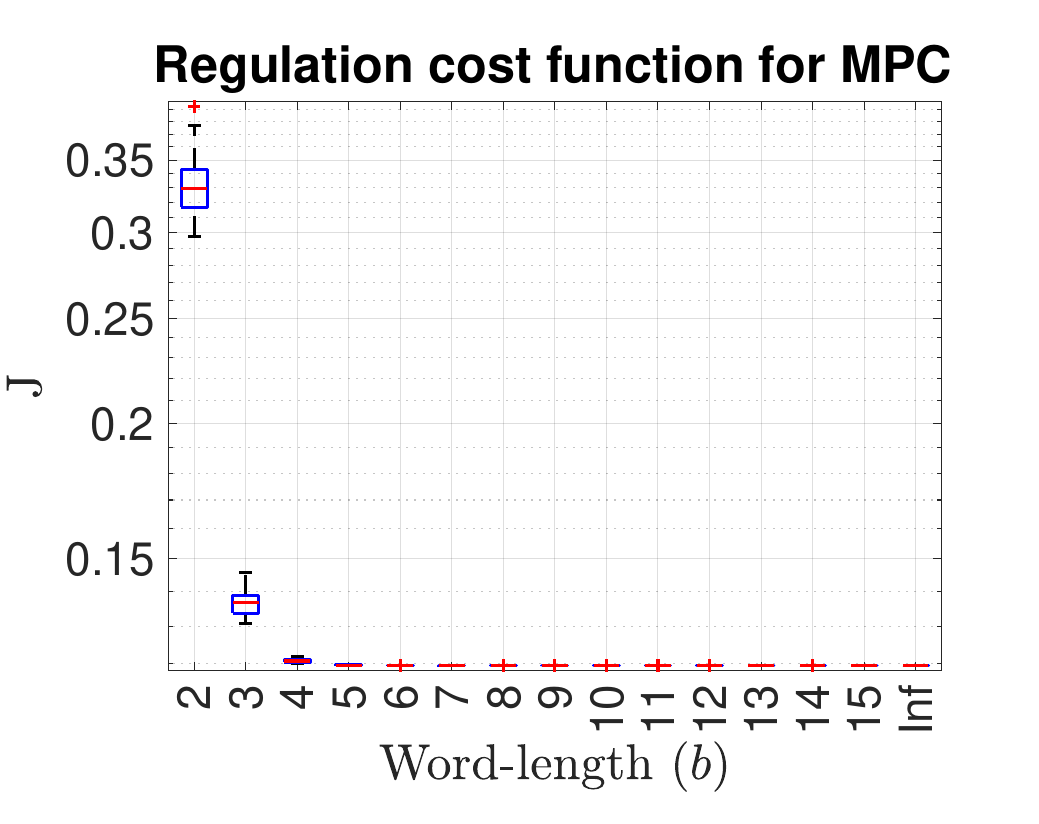}}\\
\subfloat[]{\includegraphics[trim=2cm 7cm 2cm 7cm, clip=true, width=0.33\textwidth]{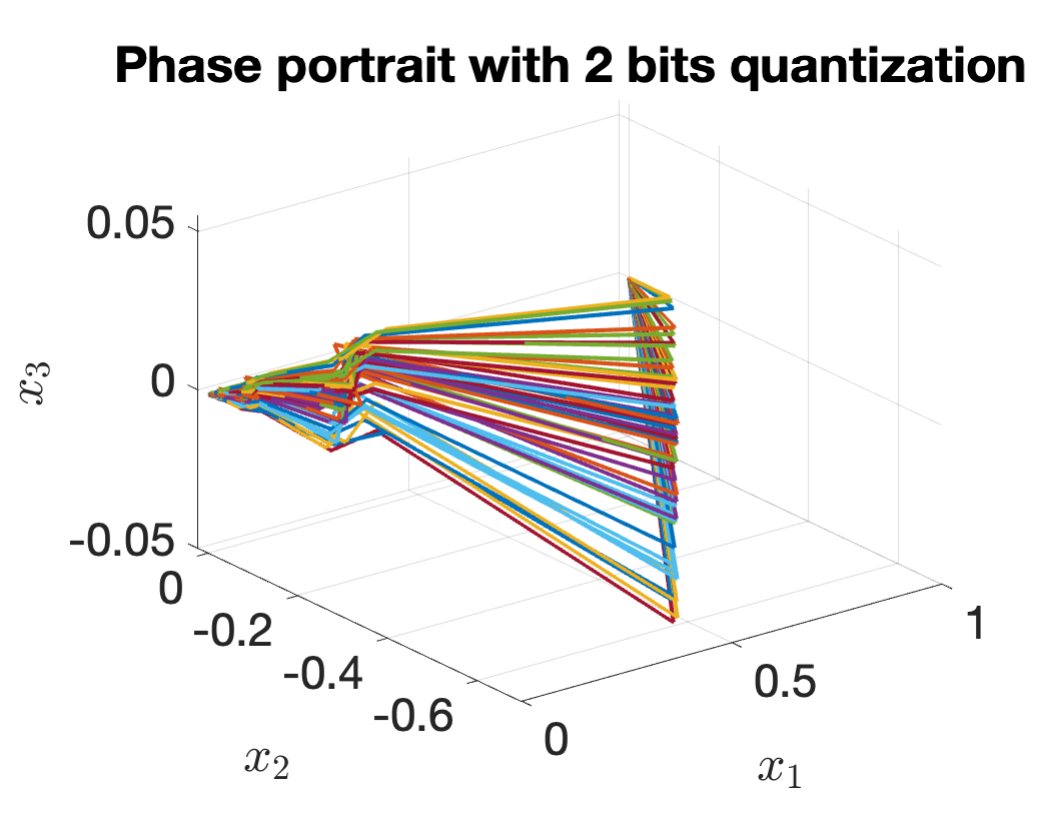}}
\subfloat[]{\includegraphics[trim=2cm 7cm 2cm 7cm, clip=true, width=0.33\textwidth]{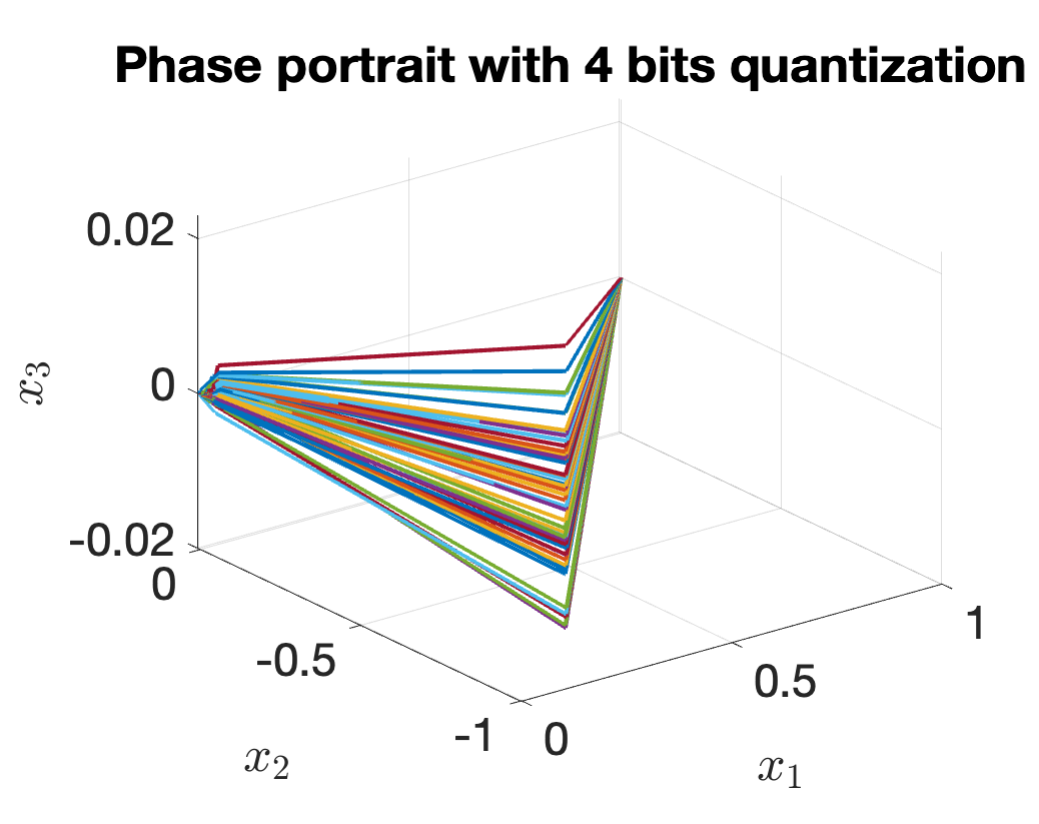}}
\subfloat[]{\includegraphics[trim=2cm 7cm 2cm 7cm, clip=true, width=0.33\textwidth]{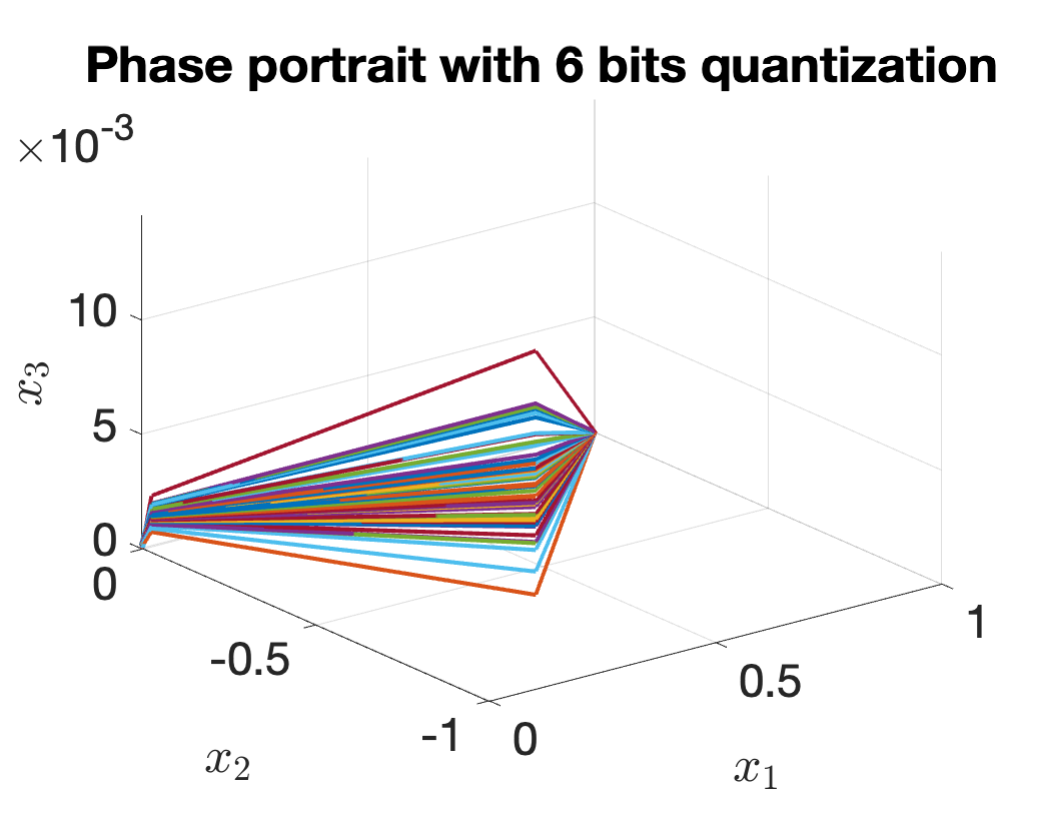}}
\caption{Error and phase-portrait profile for DC Motor with Load \eqref{eq: Motor}: (a) relative error in matrix $A$; (b) relative error in matrix $B$; (c) optimal cost achieved by MPC; (d)--(f) phase portrait from regulation MPC (with model identified from data snapshots quantized by 50 independent dither signal realization) for word lengths $b=2,\ 4, \ 6$ respectively.}
\label{Fig: Motor} \vspace{-0.2cm}
\end{figure*}
\subsection{DC Motor with Load}
A DC Motor with control torque on the load is considered for this example. The dynamics of the system are as follows:
\begin{align}\label{eq: Motor}
\begin{split}
    \dot{x}_1 &= x_2. \\
\dot{x}_2 &= \frac{K\,x_3 - c\,x_2 + u_2}{J}, \\
\dot{x}_3 &= \frac{u_1 - r\,x_3 - K_e\,x_2}{L}.
\end{split}
\end{align}
We use the following values of the motor parameters: $J = 0.01$, $c = 0.1$, $K = 0.01$, $K_e = 0.01$, $r = 1$, $L = 0.5$. Substituting these parameters and time-discretizing with a discretization interval of 1 second yields:
\begin{align*}
    A=\left[
    \begin{array}{ccc}
        1.000 & 0.099 &  0.041\\
        0 & 0 & 0.016 \\
        0 & 0 & 0.135
    \end{array}
    \right],~~~
    B= \left[
    \begin{array}{cc}
        0.048 & ~~8.996 \\
        0.083 & ~~9.991  \\
        0.864 & -0.083
    \end{array}
    \right].
\end{align*}
For the training phase, the initial conditions are generated randomly with uniform distribution in the unit cube $[-1, 1]^3$. The control input for each trajectory is chosen to be a uniformly distributed random signal on $[-1, 1]^2$. The system is simulated for 200 trajectories over 1000 sampling periods (i.e., 10 seconds per trajectory). Relative 2-norm error $\frac{\| A - \hat{A}\|}{\|A\|}$ and $\frac{\| B - \hat{B}\|}{\|B\|}$ for different word-length are shown in Fig.~\ref{Fig: Motor}(a)-(b). The identified $[\hat{A},\ \hat{B}]$ is then used to drive $x_0=[1 \ 0\ 0]^\top$ to origin by solving the MPC problem \eqref{eq:LMPC}  with $Q = \operatorname{diag}([1\ 0.1\ 0.1])$, $R = \operatorname{diag}([1\ 1])$. The minimum MPC costs that are achieved for different word-lengths are demonstrated in Fig.~\ref{Fig: Motor}(c). Fig.~\ref{Fig: Motor}(d)--(f) show the phase-portrait of controlled state-trajectories for different word-lengths. Notice that the logarithmic errors in $A$ and $B$ matrices decrease linearly with the word-length $b$ with a slope of $-0.301$ and $-0.322$. Note that, for a finite-data regime, the error should be $O(\epsilon) \approx k\epsilon = k(x_{\max}-x_{\min})/2^b$ for some constant $k$ as $\epsilon \rightarrow 0$, i.e., the logarithm of the error should decrease linearly with $b$ at a slope of $-\log 2 = -0.301$. It is also evident that the MPC Cost decreases, and the trajectories converge better with higher word-length $b$, i.e., lower quantization resolution $\epsilon$, as predicted analytically.

\begin{figure*}[t]
\centering 
\subfloat[]{\includegraphics[trim=1.9cm 7cm 2cm 7cm, clip=true, width=0.33\textwidth]{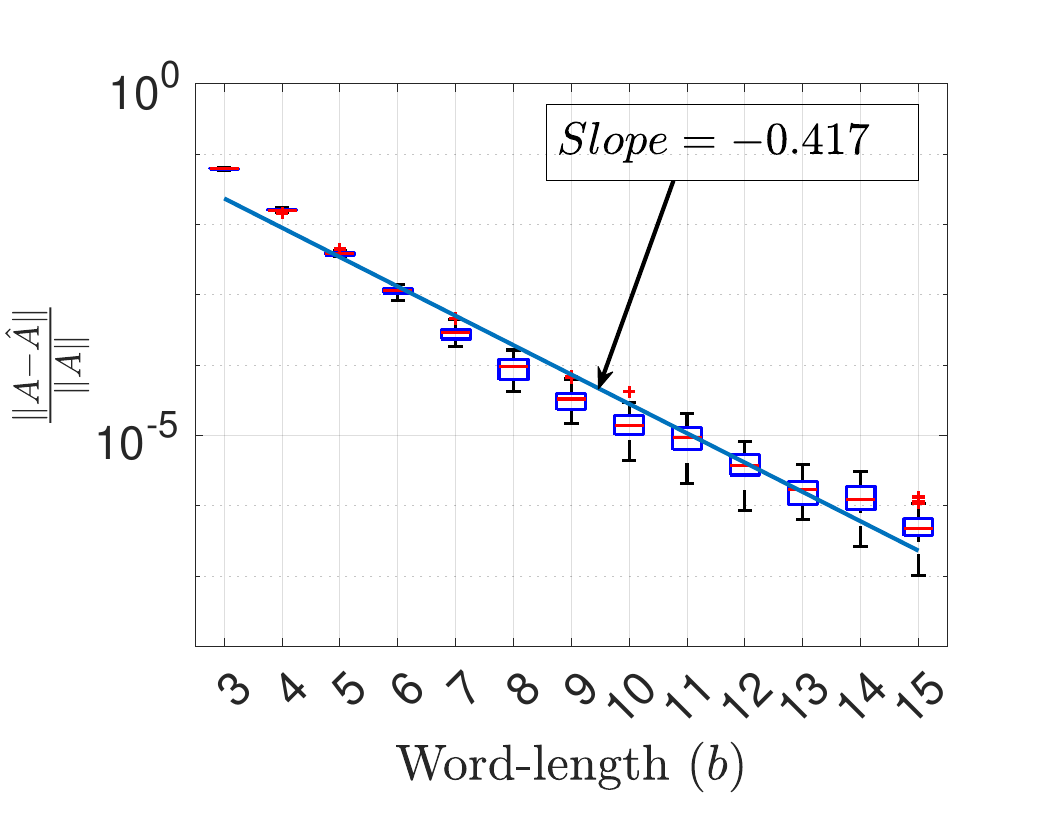}}
\subfloat[]{\includegraphics[trim=1.9cm 7cm 2cm 7cm, clip=true, width=0.33\textwidth]{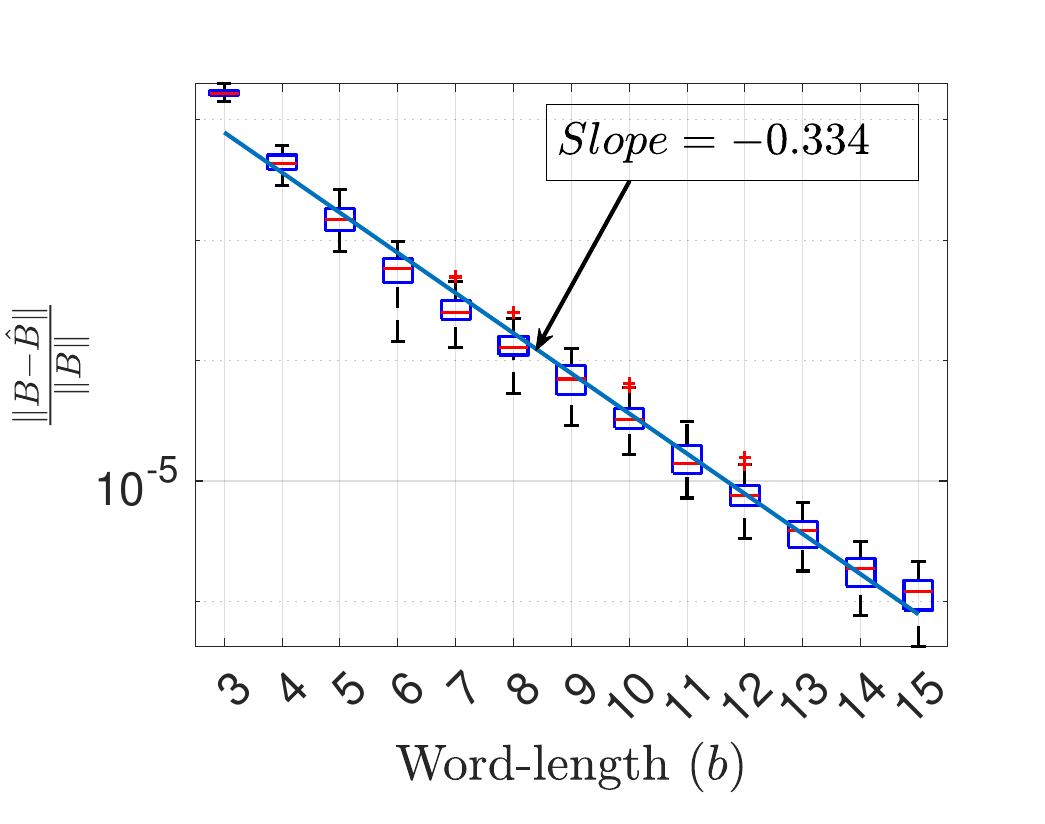}}
\subfloat[]{\includegraphics[trim=2cm 7cm 2cm 7cm, clip=true, width=0.33\textwidth]{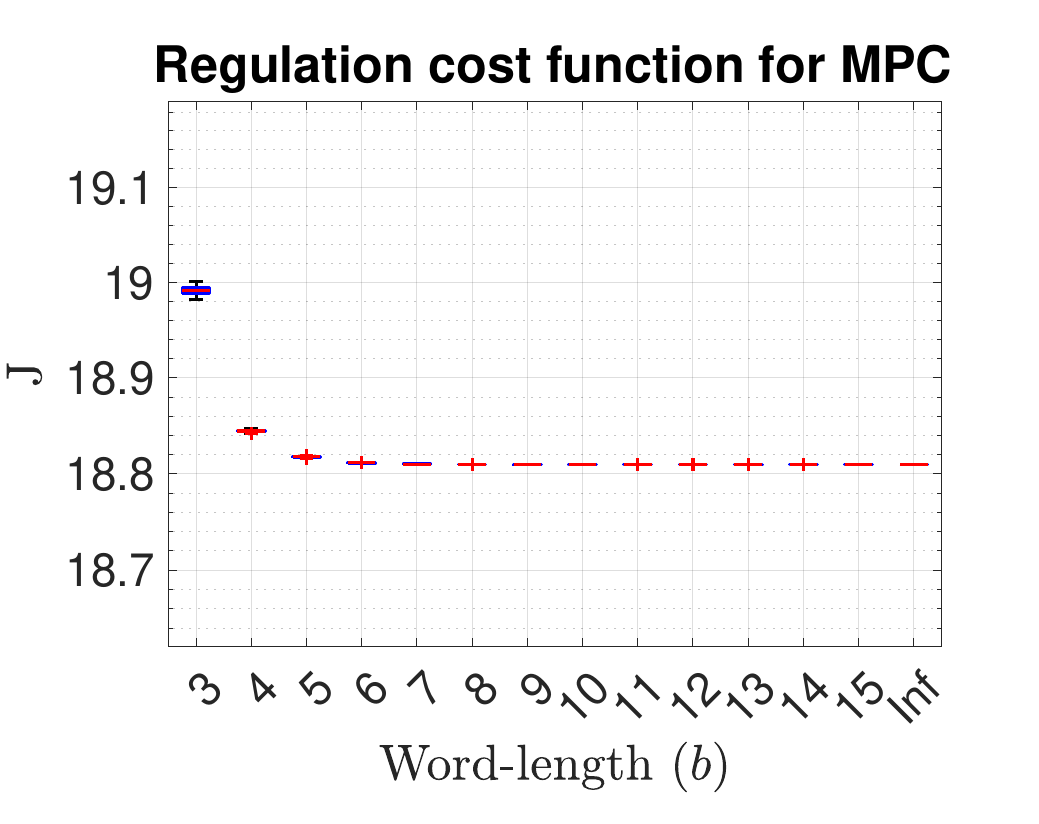}}\\
\subfloat[]{\includegraphics[trim=2cm 7cm 2cm 7cm, clip=true, width=0.33\textwidth]{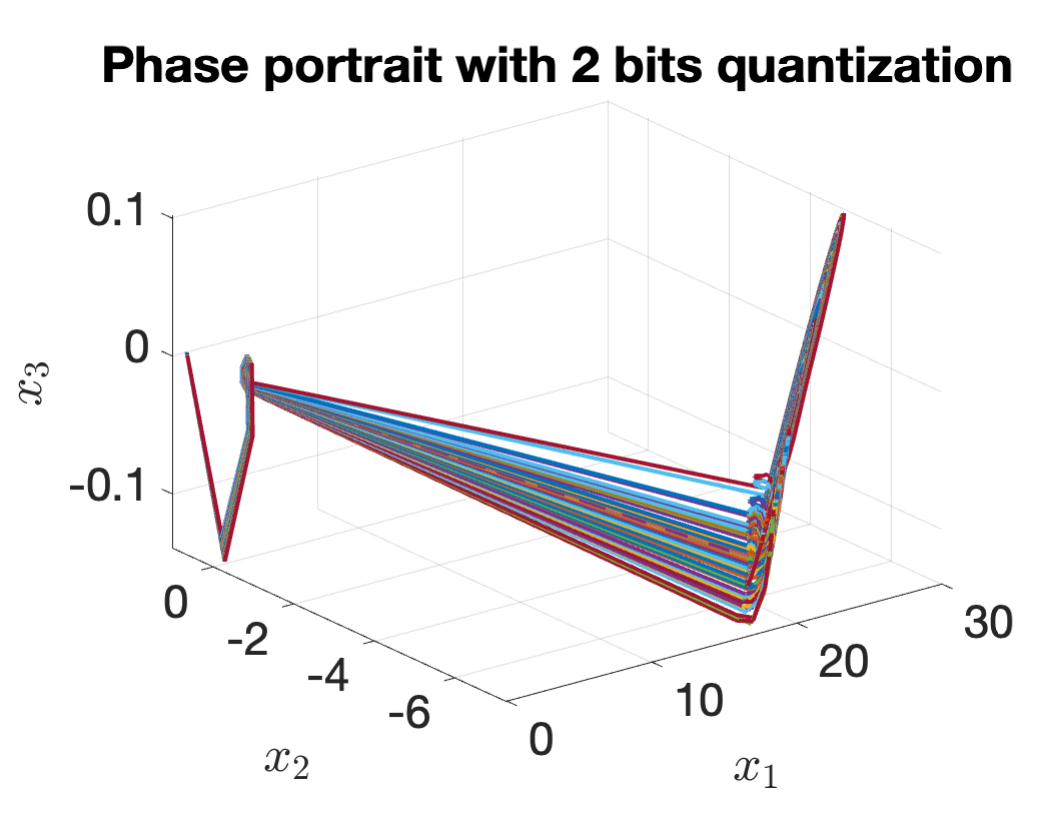}}
\subfloat[]{\includegraphics[trim=2cm 7cm 2cm 7cm, clip=true, width=0.33\textwidth]{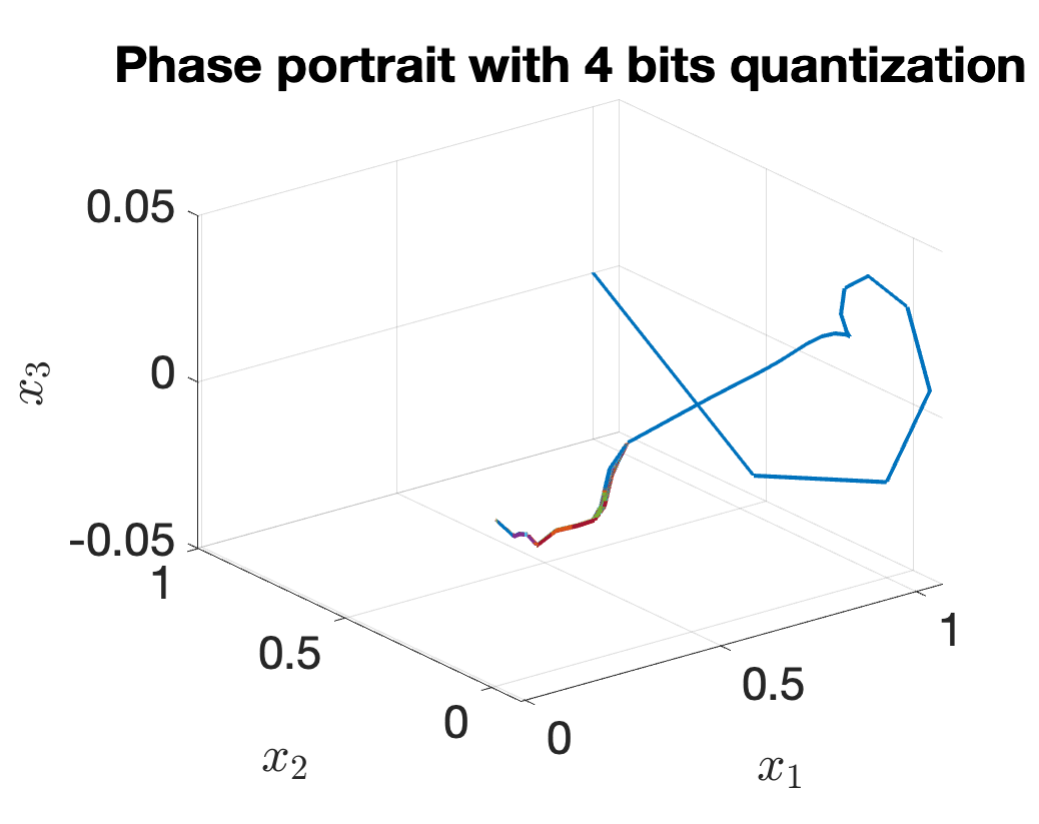}}
\subfloat[]{\includegraphics[trim=2cm 7cm 2cm 7cm, clip=true, width=0.33\textwidth]{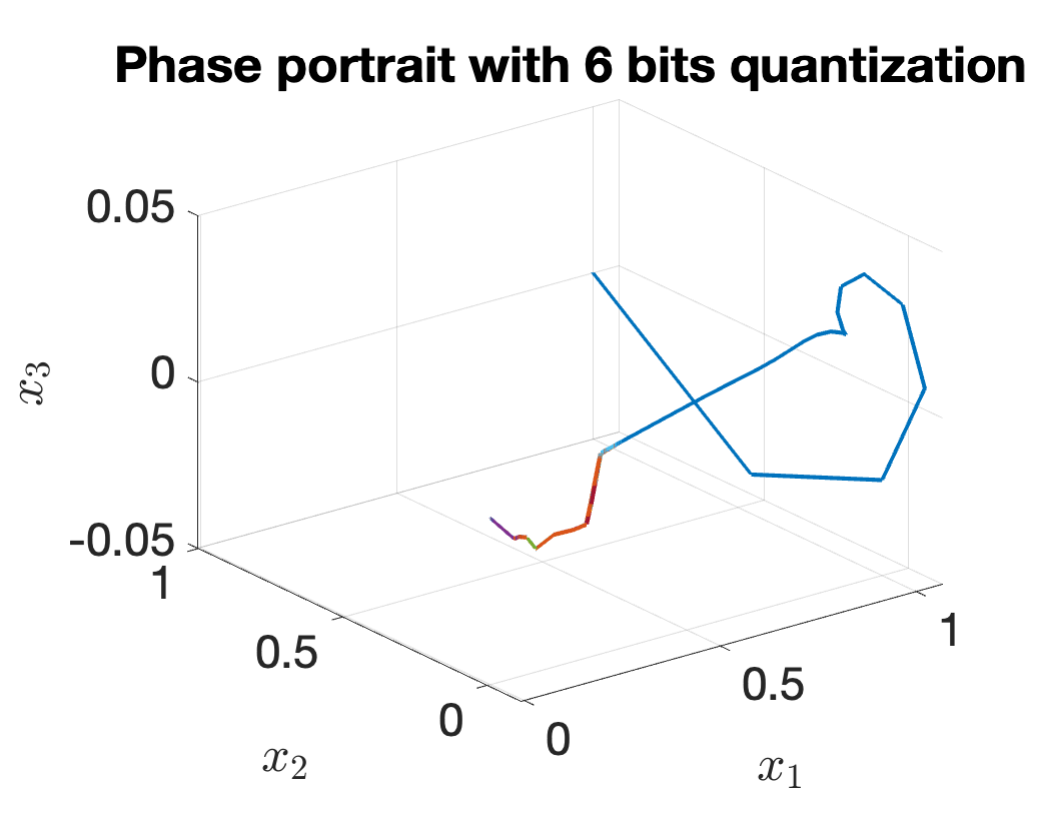}}
\caption{Error and phase-portrait profile for Boeing 747 longitudinal flight control: (a) relative error in matrix $A$; (b) relative error in matrix $B$; (c) optimal cost achieved by MPC; (d)--(f) phase portrait from regulation MPC (with model identified from data snapshots quantized by 50 independent dither signal realization) for word lengths $b=2,\ 4, \ 6$ respectively.}
\label{Fig: Boeing} \vspace{-0.2cm}
\end{figure*}

\subsection{Boeing 747 Flight Control}
For the second example, the longitudinal flight control system of a Boeing 747 is examined under linearized dynamics. Assuming steady-level flight at an altitude of 40,000 feet and a velocity of 774 feet per second, with a discretization interval of 1 second, the system matrices of discretized dynamics \eqref{eq:dynamics} are given by \cite{goel2021competitivecontrol}:

\begin{align*}
    A=\left[\!
    \begin{array}{cccc}
        0.99 & ~~0.03 & -0.02 & -0.32 \\
        0.01 & ~~0.47 & ~~4.70 & ~~0.00 \\ 
        0.02 & -0.06 & ~~0.40 & ~~0.00 \\ 
        0.01 & -0.04 & ~~0.72 & ~~0.99
    \end{array}\!
    \right],~~
    B= \left[\!
    \begin{array}{cc}
        ~~0.01 & 0.99 \\
        -3.44 & 1.66 \\
        -0.83 & 0.44 \\
        -0.47 & 0.25
    \end{array}\!
    \right].
\end{align*}
The training settings remain the same as that of the DC motor with load. Fig.~\ref{Fig: Boeing}(a) and (b) show similar trends for errors in linear predictor matrices $A$ and $B$. The identified $[\hat{A},\ \hat{B}]$ is then used to drive $x_0=[1 \ 1 \ 0\ 0]^\top$ to origin by solving the MPC problem \eqref{eq:LMPC}  with $Q = \operatorname{diag}([1\ 0.1\ 0.1 \ 0.1])$, $R = \operatorname{diag}([1\ 1])$. The minimum MPC costs achieved for different word-lengths are demonstrated in Fig.~\ref{Fig: Motor}(c). Fig.~\ref{Fig: Boeing}(d)--(f) show the phase-portrait of controlled state-trajectories for different word-lengths. We see a similar trend in identification error, and MPC cost here as well.

\section{Conclusions} \label{sec:conclusions}
In this letter, we present \texttt{\textbf{QSID-MPC}}---a model predictive control framework that utilizes system identification from dither quantized data. We theoretically analyze the connection between the true system and the system identified from quantized data. The effect of quantization is quantified for both finite and large data regimes. Our analysis shows the quantization resolution $\epsilon$ affects the identified system as $O(\epsilon)$ in finite data regime and $O(\epsilon^2)$ in large data regime. Further analysis of the performance of MPC shows that the true system is \emph{uniformly ultimately bounded} by \texttt{\textbf{QSID-MPC}} with a bound of $O(\epsilon^2)$ for the large data regime, thereby proving its robustness against quantization. Our analysis is validated via repeated trials of experiments on multiple problems. 



\bibliographystyle{ieeetr}
\bibliography{ref1, references}

\begin{thebibliography}{10}

\bibitem{Arbabi2017}
H.~Arbabi and I.~Mezi\'{c}, ``Ergodic theory, dynamic mode decomposition, and
  computation of spectral properties of the koopman operator,'' {\em SIAM
  Journal on Applied Dynamical Systems}, vol.~16, no.~4, pp.~2096--2126, 2017.

\bibitem{Folkestad2022}
C.~Folkestad, S.~X. Wei, and J.~W. Burdick, ``Koopnet: Joint learning of
  koopman bilinear models and function dictionaries with application to
  quadrotor trajectory tracking,'' in {\em 2022 International Conference on
  Robotics and Automation (ICRA)}, pp.~1344--1350, 2022.

\bibitem{Cleary2020}
A.~Cleary, K.~Yoo, P.~Samuel, S.~George, F.~Sun, and S.~A. Israel, ``Machine
  learning on small uavs,'' in {\em 2020 IEEE Applied Imagery Pattern
  Recognition Workshop (AIPR)}, pp.~1--5, 2020.

\bibitem{nair2004stabilizability}
G.~N. Nair and R.~J. Evans, ``Stabilizability of stochastic linear systems with
  finite feedback data rates,'' {\em SIAM Journal on Control and Optimization},
  vol.~43, no.~2, pp.~413--436, 2004.

\bibitem{maity2021optimal}
D.~Maity and P.~Tsiotras, ``Optimal controller synthesis and dynamic quantizer
  switching for linear-quadratic-{G}aussian systems,'' {\em IEEE Transactions
  on Automatic Control}, vol.~67, no.~1, pp.~382--389, 2021.

\bibitem{maity2023optimal}
D.~Maity and P.~Tsiotras, ``Optimal quantizer scheduling and controller
  synthesis for partially observable linear systems,'' {\em SIAM Journal on
  Control and Optimization}, vol.~61, no.~4, pp.~2682--2707, 2023.

\bibitem{gray1993dithered}
R.~M. Gray and T.~G. Stockham, ``Dithered quantizers,'' {\em IEEE Transactions
  on Information Theory}, vol.~39, no.~3, pp.~805--812, 1993.

\bibitem{maity2024effect}
D.~Maity, D.~Goswami, and S.~Narayanan, ``On the effect of quantization on
  dynamic mode decomposition,'' in {\em 2024 IEEE 63rd Conference on Decision
  and Control (CDC)}, pp.~8778--8785, 2024.

\bibitem{maity2024EDMD}
D.~Maity and D.~Goswami, ``On the effect of quantization on extended dynamic
  mode decomposition,'' {\em arXiv preprint arXiv:2410.02803}, 2024.

\bibitem{ataei2025koopman}
S.~Ataei, D.~Maity, and D.~Goswami, ``Koopman meets limited bandwidth: Effect
  of quantization on data-driven linear prediction and control of nonlinear
  systems,'' {\em arXiv preprint arXiv:2501.07714}, 2025.

\bibitem{schmid2010}
P.~J. Schmid, ``Dynamic mode decomposition of numerical and experimental
  data,'' {\em Journal of Fluid Mechanics}, vol.~656, pp.~5--28, 2010.

\bibitem{Williams2015}
M.~O. Williams, I.~G. Kevrekidis, and C.~W. Rowley, ``{A data--driven
  approximation of the Koopman operator: extending Dynamic Mode
  Decomposition},'' {\em Journal of Nonlinear Science}, vol.~25,
  pp.~1307--1346, Dec 2015.

\bibitem{fu2024tutorial}
M.~Fu, ``A tutorial on quantized feedback control,'' {\em IEEE/CAA Journal of
  Automatica Sinica}, vol.~11, no.~1, pp.~5--17, 2024.

\bibitem{widrow1961statistical}
B.~Widrow, ``Statistical analysis of amplitude-quantized sampled-data
  systems,'' {\em Transactions of the American Institute of Electrical
  Engineers, Part II: Applications and Industry}, vol.~79, no.~6, pp.~555--568,
  1961.

\bibitem{schuchman1964dither}
L.~Schuchman, ``Dither signals and their effect on quantization noise,'' {\em
  IEEE Transactions on Communication Technology}, vol.~12, no.~4, pp.~162--165,
  1964.

\bibitem{Grune2017}
L.~Gr\"{u}ne and J.~Pannek, {\em Nonlinear Model Predictive Control}.
\newblock Springer International Publishing, 2017.

\bibitem{Cruz-Hernandez1999}
C.~Cruz-Hern\'{a}ndez, J.~Alvarez-Gallegos, and R.~Castro-Linjares, ``Stability
  of discrete nonlinear systems under novanishing pertuabations: application to
  a nonlinear model-matching problem,'' {\em IMA Journal of Mathematical
  Control and Information}, vol.~16, no.~1, pp.~23--41, 1999.

\bibitem{Khalil2015}
H.~Khalil, {\em Nonlinear Systems}.
\newblock Pearson Education International Incorporated, 2015.

\bibitem{goel2021competitivecontrol}
G.~Goel and B.~Hassibi, ``Competitive control,'' {\em IEEE Transactions on
  Automatic Control}, vol.~68, no.~9, pp.~5162--5173, 2022.

\end{thebibliography}


\end{document}